# Microgrids Coalitions for Energy Market Balancing

Viorica Chifu, Cristina Bianca Pop, Tudor Cioara, Ionut Anghel
Technical University of Cluj-Napoca, Romania
Department of Computer Science

With the integration of renewable sources in electricity distribution networks, the need to develop intelligent mechanisms for balancing the energy market has arisen. In the absence of such mechanisms, the energy market may face imbalances that can lead to power outages, financial losses or instability at the grid level. In this context, the grouping of microgrids into optimal coalitions that can absorb energy from the market during periods of surplus or supply energy to the market during periods of is a key aspect in the efficient management of distribution networks. In this article, we propose a method that identify an optimal microgrids coalition capable of addressing the dynamics of the energy market. The proposed method models the problem of identifying the optimal coalition as an optimization problem that it solves by combining a strategy inspired by cooperative game theory with a memetic algorithm. An individual is represented as a coalition of microgrids and the evolution of population of individuals over generations is assured by recombination and mutation. The fitness function is defined as the difference between the total value generated by the coalition and a penalty applied to the coalition when the energy traded by coalition exceeds the energy available/demanded on/by the energy market. The value generated by the coalition is calculated based on the profit obtained by the collation if it sells energy on the market during periods of deficit or the savings obtained by the coalition if it buys energy on the market during periods of surplus and the costs associated with the trading process. This value is divided equitably among the coalition members, according to the Shapley value, which considers the contribution of each one to the formation of collective value.

## 1. Introduction

With the acceleration of the transition to renewable sources, energy systems around the world are undergoing a transformation, adopting decentralized, dynamic and interconnected models that redefine the processes of energy production, distribution and consumption. In this context, microgrids are local autonomous systems, capable of producing, consuming and storing energy [7], that play an important role in maintaining the stability of the electricity grid in the face of fluctuations or unexpected interruptions. Their importance is particularly high in urban and island areas, where reducing dependence on centralized infrastructure and integrating renewable sources are becoming strategic priorities. However, the intermittent nature of renewable production generates frequent imbalances between energy demand and supply. In the absence of an intelligent coordination mechanism, these imbalances can lead either to significant losses due to unused energy surpluses or to instability in supply during periods of high demand. In this context, the development of cooperation mechanisms between microgrids and the energy market, which are dynamically correlated with market price signals and the state of charge of the microgrid batteries, is necessary. A low price on the market signals an energy surplus of energy in the grid, while a high price indicates a deficit of available energy. Interpreting these price signals allows microgrids to make energy-coordinated decisions based on available storage capacity in battery: to store available energy when it is cheap or to supply it to the energy market when the market is in deficit and pays better for the energy purchased.

In this context, we propose a method that identifies the optimal coalition of microgrids that can respond to market imbalances either by absorbing surplus energy or by compensating for a market deficit. The decision to form a coalition is directly influenced by the price signal from the market. When the energy price increases, signaling a deficit of energy, microgrids with surplus energy are encouraged to collaborate and sell energy, thus maximizing collective revenues. Conversely, when the price decreases, indicating an energy surplus, microgrids form coalitions to buy and store energy at low costs, preparing for periods of increased demand. In our approach, we consider that each microgrid is equipped with a storage battery, whose internal state is characterized by four factors: the amount of available energy, the number of remaining usage cycles, the maximum storage capacity, and the current operational state (i.e. charged, partially charged, or discharged). These attributes define the flexibility of each microgrid to participate in energy coalitions. By integrating this information with the state of the energy market at a given time (i.e. energy deficit or energy surplus), our method allows the identification of optimal coalition, capable of responding to energy market dynamic. The identification of optimal coalition is achieved through an approach that combines cooperative game theory with a memetic algorithm. The distribution of profit obtained from energy transactions between coalition members is based on the Shapley value, which considers the individual contribution of each microgrid to the total value generated. Shapley value is calculated considering the amount of energy traded by each coalition member, as well as the costs associated with the storage batteries.

The paper is structured as follows. Section 2 presents the related work. Section 3 presents the proposed method for identifying the optimal coalition, which combines strategies inspired by cooperative game theory with a memetic algorithm. Section 4 presents the performance evaluation of the proposed method in terms of the quality of the created microgrids coalition, while section 5 presents a discussion on the performance of the memetic algorithm in terms of fitness, diversity and stability. We end our paper with conclusions and future work.

## 2. Related work

In the field of P2P energy trading [1] propose a framework for Peer-to-Peer energy trading between consumers based on a coalition hunger game. In the trading process, prosumers decide opportunistically whether or not to use their battery in the trading process in order to maximize their benefits. Each prosumer can choose between two states at the time of trading, namely whether to use the battery to buy or sell energy on the market or whether to not use the battery and sell only the surplus energy. The decision to choose between the two states is made depending on the benefits they obtain. The coalition formation game allows the formation of trading clusters in which prosumers with similar behaviors are grouped, and decisions to form or dissolve coalitions are made based on a Pareto-like merging and splitting rule. Although the results obtained show that the method is capable of forming stable coalitions and that it brings greater benefits to prosumers compared to traditional schemes (i.e. Feed-in Tariff), it has the disadvantage of being costly in terms of time and resources.

Tushar et.al 2019 [2] proposes a P2P trading framework between prosumers that combines blockchain technology with a coalition-building game and a motivational psychological framework. Blockchain technology is integrated to ensure transparency and security of transactions, and the coalition-building game is used to group prosumers into trading coalitions. The goal of coalition formation is to maximize the benefits obtained by prosumers compared to the situation in which they would participate individually in the trading process. To stimulate prosumers to participate in the transaction process, psychological models are used (i.e. Rational Economic Model, ELM – Elaboration Likelihood Model, Positive Reinforcement Model).Although the proposed model has demonstrated superior performance compared to individual trading, it may experience delays in transaction validation, is not scalable in a scenario with thousands of producers, and involves a form of indirect centralization induced by the need to validate and record transactions in the blockchain by an Energy Service Provider.

Ali et al. [3] propose a hybrid approach that combines a strategy based on cooperative game theory with a local optimization algorithm for P2P transactions. Game theory is used to create stable coalitions of heterogeneous prosumers that are differentiated by different production, consumption and storage capacities. The creation of the coalitions is done by applying merge-and-split rules and their stability is evaluated using Nash equilibrium and Pareto efficiency. After the formation of the coalitions, an optimization algorithm is used that considers real constraints (e.g. battery capacity, demand, etc.) to assign the volumes of energy traded between coalition members and optimize the distribution of benefits among them. Although the method demonstrates the benefits of prosumers' participation in the coalition, by reducing transaction costs and using the balanced energy produced and consumed, it does not model the strategic behavior of prosumers, assuming that they cooperate sincerely within the coalition.

Laayati et. al [4] proposes a P2P trading system that combines a hybrid VCG-PSO algorithm with blockchain technology. The VCG-PSO algorithm combines the Vickrey-Clarke-Groves (VCG) mechanism from game theory with the Particle Swarm Optimization (PSO) algorithm. The Vickrey-Clarke-Groves (VCG) mechanism is used to guarantee that each prosumer is rewarded according to his marginal impact on the collective welfare of the coalition, and PSO algorithm is used to identify the best combination of prices and quantities of energy traded between microgrids, such that the social welfare to be maximized. Blockchain technology is used to implement a mechanism for automatic execution of P2P transactions through smart contracts. Although transactions have been optimized to ensure a balance between energy demand and supply, the introduction of additional costs related to securing transactions on the blockchain, as well as high latency, represents a drawback.

Krayem et al. [5] propose a P2P energy trading method that combines cooperative game theory with blockchain technology. Cooperative games are used to form coalitions of prosumers that trade energy among themselves in order to ensure energy stability at the coalition level. The transaction price is negotiated at the coalition level. The Shapley value is used to calculate the benefit of each participant in the coalition depending on its contribution. Blockchain technology is used to record and validate energy transactions between prosumers without the need for an intermediary. Although the results have shown that coalition members obtain higher benefits than if they transacted with a centralized network, the proposed method faces scalability problems due to the use of blockchain technology.

Similar to [5], [6] proposes a game-theoretic approach to creating trading coalitions. A trading scheme is modeled using a Stackelberg game model, and the goal is to optimize the allocation of energy resources between participants, reduce the costs of consuming microgrids, and increase the profits of producing microgrids. Transactions are dictated by producers. Each producer sets the price at which they are willing to see energy, and the consumer can decide whether to trade with a particular producer or not by comparing the price with that set by the main grid. The results showed that microgrids participating in coalitions obtain a higher profit than those that would have acted individually, and the Shapley value ensures the correct distribution of this profit, depending on the contribution of each microgrid.

Unlike the existing state-of-the-art approaches in which energy exchanges are made only between local participants, and do not take into account the energy sold or bought from the market and the fact that market prices fluctuate depending on supply and demand, in this paper we propose a method that identifies coalitions of microgrids that trade directly with the energy market, maximizing the profit obtained from collation in deficit periods and minimizing the costs associated with the acquisition of energy in surplus periods. The use of a heuristic algorithm ensures efficient exploration of the solution space, and the use of the Shapley value warrants the correct distribution of benefits among coalition members.

## 3. Shapley value-based microgrid coalition formation for energy market balancing

We consider a local community of microgrids producing renewable energy:

$$MGC = \{MG_i | i = \overline{1,..,n}\} \quad (1)$$

where *n* is the number of microgrids in the community.

At a time *t*, it is necessary to identify an optimal coalition that can satisfy the energy demand or surplus in the energy market. The decision of the community is to form coalitions according to the state of the energy market. If there is a deficit in the market and the price of energy is high because of this deficit, the coalitions will decide to sell the surplus energy to maximize profits. Otherwise, if there is a surplus of energy in the market and the market price is low, the coalitions will decide to buy energy to increase the energy level in the batteries. Each microgrid in the community is equipped with a battery. The state of a microgrid at time *t* is determined by the state of its battery which is defined as follows:

$$State(ESB, t) = < E_{store}(t), noCycles(t), E_{max}, status(t) > \quad (2)$$

where: $E_{store}(t)$ is the amount of energy stored in the battery, $noCycles(t)$ is the number of remaining charge/discharge cycles, $E_{max}$ is the maximum amount of energy of battery that can be stored in battery and $status(t)$ is the current state of the battery defined as:

$$status(t) = \begin{cases} fully\ charged, if\ E_{store}(t) = E_{max} \\ partially\ charged, if\ 0 < E_{store}(t) < E_{max} \\ discharge, if\ E_{store}(t) = 0 \end{cases} \quad (3)$$

The estimated quantity of energy that can be stored in battery at a specific time moment, *t* is defined as [8]:

$$E_{store}(t) = \begin{cases} E_{store}(t-1) + \eta_{charge} * \Delta E(t), if\ \Delta E(t) > 0 \land \Delta E(t) < E_{store}(t) < E_{max} \\ E_{store}(t-1) - \frac{1}{\eta_{discharge}} * |\Delta E(t)|, if\ \Delta E(t) < 0 \land \Delta E(t) < E_{store}(t-1) \end{cases} \quad (4)$$

where $\Delta E_i(t) = E_{generated}(t) - E_{consumed}(t)$ is the difference between produced and consumed energy at the microgrid level, $\eta_{charge}$ is the charge efficiency of the battery, $\eta_{discharge}$ is the battery discharge efficiency and $E_{max}$ is the maximum quantity of energy that can be stored in the battery.

We denote with *EM* the energy marketplace. The state of the energy marketplace at time *t* is defined as follows:

$$State(EM, t) = < E_{EM}, Price_{EM}, status_{EM} > \quad (5)$$

where $E_{EM}$ is the energy surplus /deficit on the market at time *t*, $Price$ is the sale/purchase price of energy on the market at time *t*, and $status_{EM}$ in the status of energy marketplace at time *t*.

$$status = \begin{cases} deficit, if\ E_{EM} < 0 \\ surplus, if\ E_{EM} > 0 \end{cases} \quad (6)$$

Our goal is to identify the optimal microgrid coalition, $C_{opt} \subseteq MGC$ that buys energy at lower prices available in the market when there is an energy surplus on the market or sells energy to the market at a higher price when there is an energy deficit on the market. To identify the optimal coalition, we have used an approach based on cooperative game theory. We denote with *CS* the set of all possible coalitions of the set of microgrids:

$$CS = \{ C_i \sqsubseteq MGC | C_i \neq \emptyset \} \quad (7)$$
$$C_i \cap C_j = \emptyset, \forall i \neq j \quad (8)$$

where MGC is the local microgrids community. In our approach, a microgrid coalition must satisfy the following conditions:

$$if \ (status(EM) = deficit) \Rightarrow \sum_{MG_i \in C} E_{store}^{MG_i}(t) \cong E_{EM} \quad (9)$$
$$if \ (status(EM) = surplus) \Rightarrow \sum_{MG_i \in C} (E_{max}^{MG_i} - E_{store}^{MG_i}(t)) \cong E_{EM} \quad (10)$$

The characteristic function $v(C)$ of a coalition $C$ measures the benefits that coalition can bring to its members. It reflects both the revenues or savings generated from energy transactions, as well as the costs associated with the transactions. In our case, the characteristic function is defined based on the profit obtained from selling energy at higher price or the savings obtained by buying energy at lower price and the costs due to the transaction process (e.g. battery degradation cost, operating cost, etc.).

$$v(C,t) = \begin{cases} Price * \min(\sum_{MG_i \in C} E_{store}^{MG_i}(t), E_{EM}(t)) - \sum_{MG_i \in C} cost_{ESB}^{MG_i}(t) - cost_{operating}(C), if \ E_{EM} \leq 0 \\ \sum_{MG_i \in C} Price * \min(E_{max}^{MG_i} - E_{store}^{MG_i}(t)), E_{EM}) - \sum_{MG_i \in C} cost_{ESB}^{MG_i}(t) - cost_{operating}(C), if E_{EM} > 0 \end{cases}$$
(11)

where *Price* is selling/buying price of energy on the market at the time $t$, $\sum_{MG_i \in C} E_{store}^{MG_i}(t)$ is the amount of energy stored in the batteries of the microgrids in the coalition $C$, $\min(E_{max}^{MG_i} - E_{store}^{MG_i}(t))$ is the total capacity available for storage in the batteries of the microgrids in the coalition $C$, $\sum_{MG_i \in C} cost_{ESB}^{MG_i}$ is the sum of cost associated with batteries degradation of microgrids in the coalition, $E_{EM}$ is the energy surplus/deficit on the market, and $cost_{operating}(C)$ is the operating cost at the coalition level.

The cost associated with battery degradation of each microgrid in coalition is defined as follows:

$$cost_{ESB}^{MG_i}(t) = \delta * NoCycles(ESB) \quad (12)$$

where: $\delta \in (0,1)$ is the coefficient that quantifies the rate of battery damaged per unit charge and discharge cycle; $NoCycles(ESB)$ is the total number of charge/discharge cycles performed so far.

The operating cost at the level of coalition is defined as:

$$cost_{operating} = cost_{maintenance} \quad (13)$$

where $cost_{maintenance}$ is the cost associated with the maintenance of microgrid equipment.

Any coalition $C$ has a non-negative value for the characteristic function. Also, for any two disjoint coalitions $C_i$ and $C_j$, the combined value of the two coalitions is at least equal to the sum of their individual values:

$$v(C_i \sqcup C_j) > v(C_i) + v(C_j) \quad (14)$$

This means that the microgrids acting as players have an incentive to cooperate because the total value obtained by the coalition is at least as large as the sum of the values obtained separately.

To distribute the value of the characteristic function among the coalition members, we used a distribution method proportional to the contribution of each microgrid, based on the Shapley value. More specifically, each microgrid receives a share of value proportional to its energy contribution to the total amount of energy traded on the market. The Shapley value for each microgrid $MG_i$ is calculated using the formula:

$$\phi_{MG_i} = \sum_{C \subseteq MGC \setminus \{MG_i\}} \frac{|C|!(|N|-|C|-1)!}{|N|!} - (v(C \cup MG_i) - v(C)) \quad (15)$$

where: *MGC* is the set of all microgrid, $C$ is a subset of *MGC* which does not include the $MG_i$, $v(C)$ is the value of the coalition $C$.

Our goal is to identify the coalition that maximizes the total value and distributes the profit fairly among the microgrids in the coalition. In the context of energy trading, this means identifying the optimal coalition of microgrids that, in a scenario with an energy surplus on the market, is able to absorb the excess energy, and in the case of an energy deficit on the market, is able to provide the energy necessary to restore balance. At the same time, it must ensure the fair and proportional distribution of the benefits obtained from cooperation between the coalition members, depending on the contribution of each microgrid to the total volume of energy traded.

$$C = \max_{C \subseteq CS} v(C), \forall MG_i \in C, value(MG_i, v(C)) = \phi_{MG_i} \tag{16}$$

To identify the optimal coalition that can satisfy the energy surplus or deficit in the market, we used a memetic algorithm [9]. Memetic algorithms are metaheuristic algorithms that combine genetic algorithms with local optimization techniques (e.g. Simulated Annealing) to improve algorithm convergence. To apply the memetic algorithm in the context of identifying the optimal coalition of microgrids, we need to define the representation of an individual and the fitness function for evaluating the quality of the individual and develop a strategy for generating the initial population of individuals. In our case, an individual is represented as a possible coalition of microgrids in our community. To maintain a fixed size for all individuals, we represent an individual as:

$$indiv = C = \{(MG_i, flag_i) | MG_i \in MGC\}, lenght(indiv) = |MGC| \tag{17}$$

where $flag_i$ is an activation status that specifies if the microgrid $MG_i$ is part of the coalition or not (i.e. 1 if the microgrid is part of the coalition and zero otherwise).

The fitness function is defined as follows:

$$fitness(indiv) = v_{normalized}(C) - penalty \tag{18}$$

where $v_{normalized}(C)$ is the normalized value of the characteristic function, which represents the total value generated by the coalition $C$, and penalty is a penalty factor that penalizes an individual if the energy supplied or requested by the coalition represented by that individual is higher or lower than the energy requested or supplied by the energy market. The normalized value of the characteristic function is defined as:

$$v_{normalized}(C) = \frac{v(C) - \mu}{\sigma} \tag{19}$$

In this formula, $\mu$ is the average of the values of the characteristic functions and $\sigma$ is the standard deviation of these characteristic function values across the entire population of coalitions.

The penalty factor is defined as:

$$penalty = \begin{cases} 0, & if\ E_{EM} = E_c \\ \rho * |E_{EM} - E_c|, & otherwise \end{cases} \tag{20}$$

where: $\rho$ is a penalty factor that determines how severe the penalty is for an energy imbalance; $E_{EM}$ is the requested energy of the market or the market demand for energy; $E_c$ is the amount of total energy produced or consumed by the coalition $C$.

The initial population of individuals was randomly generated. For each individual in the initial population, we set a percentage $p$ (i.e. 10%) of the microgrids that are active using a probabilistic approach. For each microgrid we generated a value r ∈ [0,1] and based on it we set the flag using the following formula:

$$flag = \begin{cases} 1, & if\ r < p/100 \\ 0, & otherwise \end{cases} \tag{21}$$

The memetic algorithm (see Algorithm 1) used for identifying the optimal coalition of microgrids consists of four steps. In the first step, the initial population of individuals (i.e. coalitions of microgrids) is randomly generated, while in the second step the fitness function for each individual is computed. In the third step the genetic algorithm is applied to the initial population. In the process of applying the genetic algorithm to the initial population, rank-based selection is used to select the parents on which the crossover and mutation are applied. For the crossover we have applied a two-points crossover operator which combines two parents by selecting two crossover points and exchanging the segments between these points to create two new offspring. For the mutation, we have implemented a one-point mutation operator that uses a probabilistic threshold to decide whether to activate/deactivate a microgrid in the coalition, given the energy surplus or deficit and the associated costs. The mutation operator is applied as follows:
- Randomly select a microgrid from the coalition (i.e. individual).
- For the selected microgrid, compute the activation/deactivation probability using the formula:

$$P_{activation}(flag) = \begin{cases} \max\left(0, \min\left(1, \frac{E_{store} - cost}{E_{EM}}\right)\right), & for\ selleing \\ \max\left(0, \min\left(1, \frac{E_{max} - E_{store}}{E_{EM}}\right)\right), & for\ bying \end{cases} \tag{22}$$

where $E_{store}$ is the current energy stored in the microgrid battery; $E_{max}$ is the maximum amount of energy that can be stored in the battery, and $E_{EM}$ is the energy surplus/shortage on the market.
- Compare the activation probability with a random number generated between 0 and 1. If the activation probability is higher than the random number, the microgrid is activated in the coalition (the flag becomes 1). Otherwise, the microgrid remains inactive (the flag becomes 0).

In the final step, simulated annealing [10] is applied to a subpopulation containing the best individuals selected based on fitness value. For each individual in the subpopulation, a neighboring individual is generated by applying a small perturbation to it (i.e. by activating or deactivating a microgrid, depending on the energy surplus or deficit on the market). Then, the neighbor's solution is accepted based on an acceptance probability:

$$P_{SA} = \begin{cases} 1, & if\ \Delta F \leq 0 \\ e^{-\frac{\Delta F}{T}}, & otherwise \end{cases} \quad (23)$$

where $T$ is the current cooling temperature and $\Delta F = fitness(indiv_{new}) - fitness(indiv)$. The cooling temperature is updated according to the following formula: $T = \alpha * T$ where $\alpha \in [0,1]$ is a cooling factor.

The algorithm continues to iterate through the above steps until a stopping condition is met, such as reaching a maximum number of iterations. After completing the iterations, the memetic algorithm with Simulated Annealing returns the best individual in the population, which represents an optimal coalition of microgrids.

---

**ALGORITHM 1: Mnemonic Algorithm for energy trading of local renewable energy communities in Energy Markets**

**Inputs:** MGC – microgrids community, nolter - maximum number of iterations, popSize - population size, percentage - percentage of the microgrids that will have the flag initially set to 1, *p* - selection probability in genetic algorithm, *k* – individuals number selected for applying Simulated Annealing, $T_0$- initial temperature

**Outputs:** $indiv_{best}$ — best individual in population

**Comments:**

**Begin**
1. Population = GENERATE_INITIAL_POPULATION (MGC, *popSize*)
2. Foreach *indiv* in Population do
3.   ACTIVATE_FLAGS(indiv, percentage)
4. i = 1
5. **while** (*i ≤ nolter*) do
6.   T = INITIALIZE($T_0$)
7.   Foreach *indiv* in *Population* do
8.     Fitness(*indiv*) = COMPUTE_FITNESS (*indiv*)
9.     Fitness (Population)= Fitness (Population) ∪ fitness(indiv)
10.   End Foreach
11.   Population = RANK (Population, Fitness (Population))
12.   $parent_1$= RANK_SELECTION (Population, *p*)
13.   $parent_2$= RANK_SELECTION(Population, *p*)
14.   OffSprings = CROSSOVER ($parent_1$, $parent_2$)
15.   Foreach offspring in OffSprings do
16.     Fitness (Offsprings)= ∅
17.     offspring = MUTATE (offspring)
18.     Fitness(offspring) = COMPUTE_FITNESS (offspring)
19.     Fitness (Offsprings)= Fitness (Offsprings)∪ fitness(offspring)
20.   Endfor
21.   Population= Update (Population, Offsprings, Fitness(offsprings))
22.   Subpopulation = SELECT_BEST (POPULATION, *k*, Fitness (Population))
23.   Population = Population - Subpopulation
24.   Foreach indiv in Subpopulation
25.   **while**(T<$T_{min}$) do
26.     T = UPDATE(T, α)
27.     $indiv_{new}$ = NEIGHBOUR (indiv)
28.     If ($P_{SA}$(Fitness(indiv), Fitness($indiv_{new}$,T )≥ Random(0,1)
29.      **indiv**=$indiv_{new}$
30.   **endwhile**
31.   Population = Population ∪ Subpopulation
32.   i++
33. **endwhile**
34. $indiv_{best}$= SELECT_BEST(Population), Fitness(Population)
35. **return** $indiv_{best}$
**End**

## 4. Performance Evaluation

To evaluate the proposed approach, we have considered a scenario in which there is a deficit of energy on the market at a specific time moment. The data used in the evaluation were generated based on the Zenodo dataset [11], which contains information on renewable energy production, energy consumption and storage batteries of 50 prosumers, as well as buy price on the energy market. Since the data was collected at 15-minute intervals, preprocessing was necessary to aggregate them at one-hour intervals. This aggregate value allows us to evaluate the energy surplus or deficit of each microgrid during an hour. Based on the difference between the energy produced and the energy consumed in each hour, we calculated the energy surplus or deficit at the microgrid level. If the difference is positive, it means that there is an energy surplus and the microgrid will charge the battery. The energy available for charging is calculated considering the limits of the battery capacity and its charging rate. To model battery degradation, we used the Weibull distribution and considered that a new battery can have a maximum of 6000 charge cycles.

Figure 1 illustrates the energy available in the microgrid batteries before trading and the amount of energy that can still be stored in each battery, while Figure 2 shows the number of charge cycles remaining for each battery before trading.

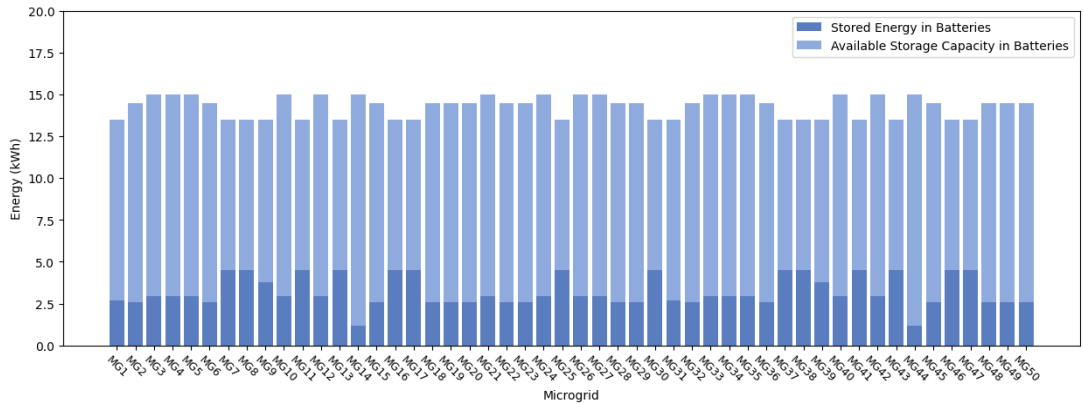

Figure 1: Initial energy stored in batteries and available storage capacity in batteries

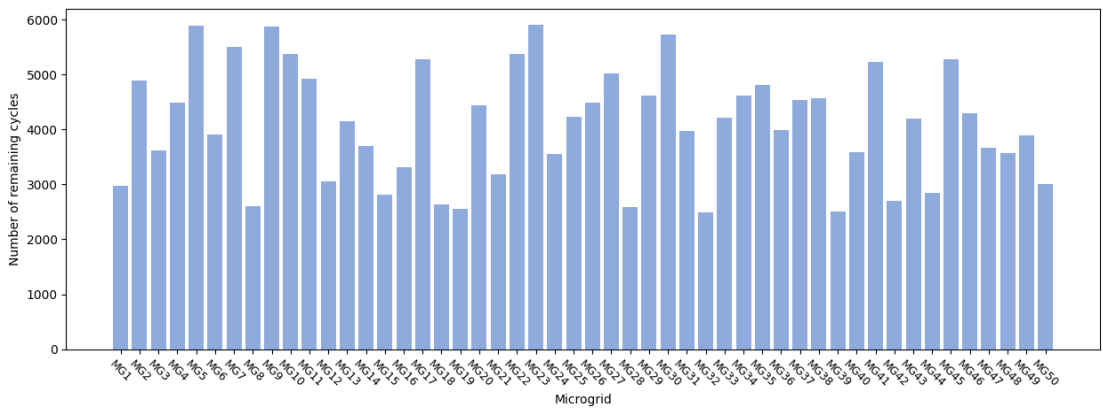

Figure 2: Initial number of charge cycles for batteries

The graph in Figure 1 shows that the microgrids analyzed have a total storage capacity ranging between approximately 12.5 and 15.5 kWh. In all cases, the energy already stored in the batteries is less than the maximum battery capacity. The distribution of stored energy levels varies from one microgrid to another, with some presenting a higher degree of charging and others a lower one. This diversity suggests that each microgrid can participate in the trading process to absorb the surplus energy on the market. Regarding the number of remaining charge cycles (See Figure 2), some batteries have a number of remaining cycles greater than 500 and others less than 3000. This indicates that microgrids with a higher number of remaining charge cycles have a higher probability of being selected to be part of the optimal coalition.

For considered scenario we analyse the following aspects: (a) the distribution of microgrids included in coalition versus excluded ones, depending on battery storage capacity; (b) distribution of remaining charge cycles for microgrids included and excluded from the coalition; (c) the evolution of the energy stored in the battery of each microgrid in the coalition during the transaction, and (d) correlation between the battery storage capacity of each microgrid, associated costs, and value distributed to each microgrid based on Shapley value. Figure 3 (a) illustrates the distribution of microgrids included in the optimal coalition versus those excluded, depending on the energy available in the battery, and Figure 3 (b) shows the distribution of the remaining number of charge cycles for each microgrid.

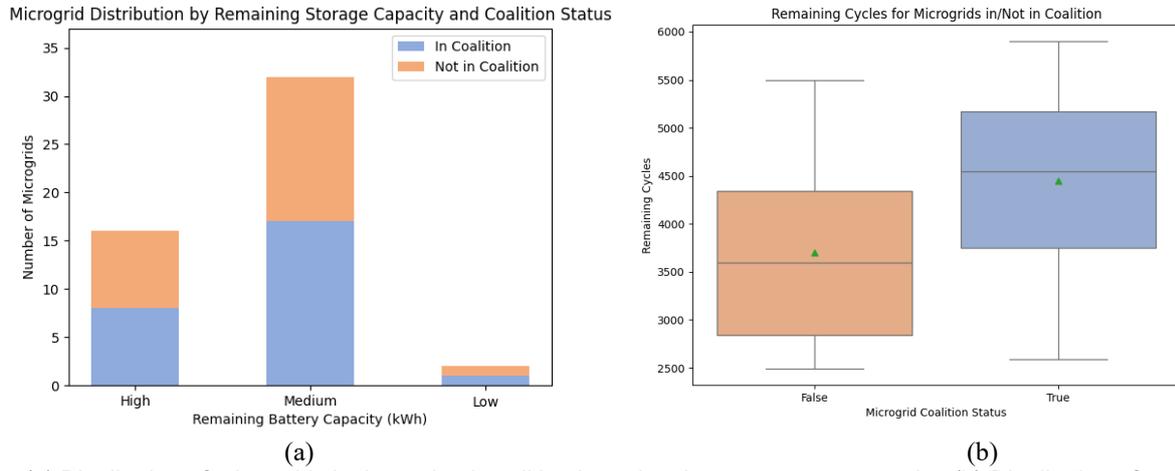

Figure 3 : (a) Distribution of microgrids in the optimal coalition based on battery storage capacity; (b) Distribution of remaining charge cycles for selected vs excluded microgrids

The two graphs from Figure3 (a) and (b), suggest that microgrids with medium or large storage capacity and with more charge cycles remaining were selected by the algorithm to be part of the optimal coalition. This behavior highlights the algorithm's ability to create a coalition that absorbs as much energy available on the market as possible, but at the same time minimizes the risks of rapid battery degradation.

The graph in Figure 4 shows the evolution of the battery charge level of the microgrids in coalition. This graph illustrates how the amount of energy in the battery of each microgrid in the coalition changes, highlighting the final energy state reached by each microgrid at the end of the trading process.

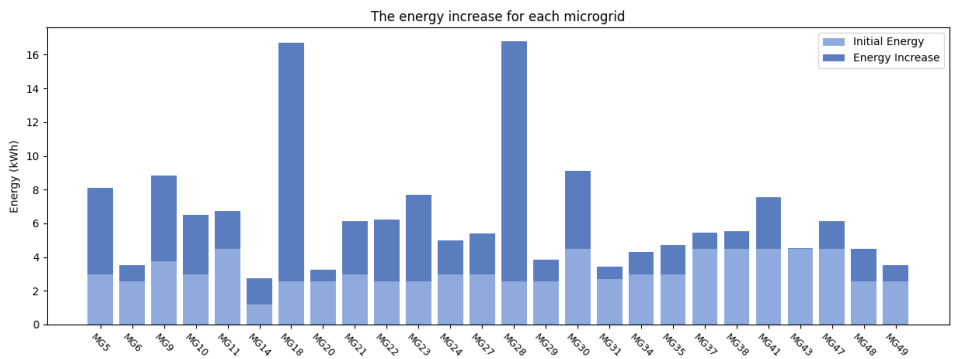

Figure 4: The level of energy storage in the batteries during trading

We observe that although the proposed algorithm identified a coalition of microgrids that is characterized by the fact that most of the participating microgrids have recorded energy increases following trading, there is still one microgrid (MG41) that, although part of the coalition, does not use the available storage capacity optimally. This suggests that the algorithm tries to ensure a compromise between maximizing the collective energy benefit and minimizing the costs due to battery degradation and maintenance, as well as penalties for exceeding the energy available in the market.

The heatmap in Figure 5 illustrates the correlation between the battery storage capacity of each microgrid from coalition, the associated costs and the value distributed by the Shapley method.

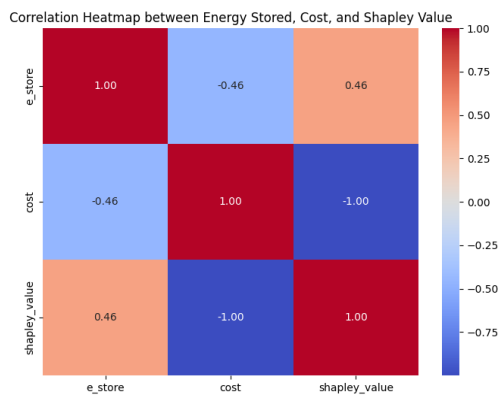

Figure 5: Heatmap for view correlation between battery energy, cost and distributed Shapley value per microgrid

The moderate positive correlation ($\simeq 0.46$) between storage capacity and Shapley value, suggests that microgrids with higher energy storage capacity are assessed as having a more valuable contribution within the coalition. This means that as the storage capacity available in microgrids increases, the Shapley value assigned to that microgrids also increases. Moreover, the perfectly negative correlation ( $\simeq$-1.0) between cost and Shapley value indicates a strict inversely proportional relationship, which demonstrates that microgrids with higher costs receive a lower value of the total Shapely value of the coalition. These results validate that our algorithm ensures an equitable distribution of benefits between coalition members.

# 5. Discussion

To evaluate the performance of the memetic algorithm and determine the optimal configuration of its adjustable parameters, we adopted a two-phase testing methodology. In the first phase, the goal was to analysis of the impact of adjustable parameters (i.e. population size, number of generations and cooling factor) on algorithm. The second phase focused on the detailed analysis of the behaviour of the algorithm for the identified optimal configuration, studying the evolution of fitness over generations, the stability between successive runs, the ability to find global optimal solutions, and the complexity of the algorithm, including the impact of parameters on execution time. This methodology was designed to ensure a rigorous evaluation of the algorithm, minimizing the effects of variability caused by random factors and ensuring a correct interpretation of the influence of each adjustable parameter on algorithm performance.

## 5.1 Analysis of the Impact of Adjustable Parameters on Algorithm Performance

To identify the optimal configuration of the adjustable parameters, we considered a methodology consisting of the following steps: (a) selection of test intervals for the adjustable parameters, (b) exploring the relationships between the adjustable parameters and the fitness value, (c) modelling the relationship between the adjustable parameters and fitness through a regression model, and (d) validating the regression model. The optimal configuration was identified by applying the methodology on the scenario, in which there was a deficit in the market, and this was covered by the microgrids coalition. In our experiments, for each individual in the initial population, we set the percentage of flags for which the value is 1 (corresponding to the microgrids initially set to participate in the trading process) to 0.1. The penalty factor, denoted by ρ, which regulates the severity of the penalty for an individual in case an energy imbalance is reached in the market during trading, was set to a value of 0.5. Also, the percentage $k$ of the population on which the simulated annealing method is applied was set to 20%. To investigate the influence of each parameter on fitness, we considered four independent variables (i.e. adjustable parameters): Population size (PopSize), number of generations (GenNumber) and Cooling Factor (CoolingFactor). For PopSize the range of variation was set to [20, 300], and variation step to 10. For GenNumber the range of variation was set to [20, 300] and the variation step to 10, while for CoolingFactor)the range of variation was set to [0.1, 1] and the variation step 0.For each possible combination of parameters, the algorithm was run 50 times and the average of the final fitness obtained was recorded to reduce the effect of randomness. To understand the relationship between adjustable algorithm parameters and fitness value, we used scatterplots and correlation coefficients. Figure 26 illustrates how each of the three adjustable parameters of the algorithm influences the fitness value, while the other two are held constant. Figure 6(a) shows how fitness varies with population size, with the number of generations set to 100 and the cooling factor set to 0.5. Figure 6(b) shows the variation of fitness with the number of generations, keeping the population size at 100 and the cooling factor set to 0.5. Figure 6 (c) highlights the influence of the cooling factor on fitness, using a population of 100 and 100 generations.

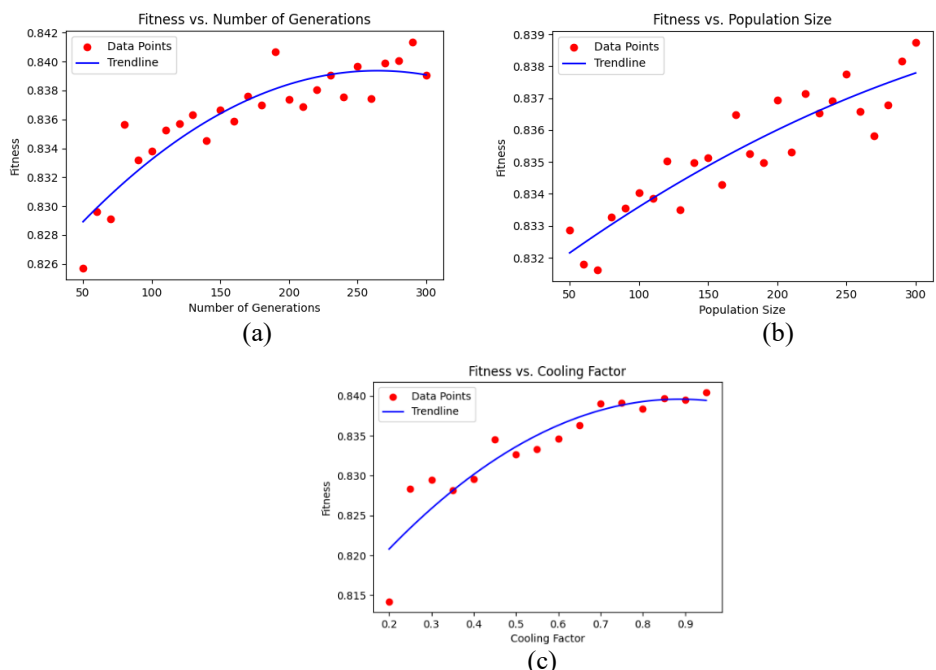

Figure 6: (a) Fitness vs GenNumber; (b) Fitness vs. PopSize; (c) Fitness Evolution vs Cooling Factor

The analysis of the scatterplots highlights that the relationship between population size and fitness is approximately linear. In contrast, the relationships between the number of generations and fitness, and the cooling factor and fitness, respectively, are nonlinear. In all three cases, an increase in the fitness value is observed as the adjustable parameter analysed increases. However, in the case of the number of generations and the cooling factor, after a certain value, the fitness stabilizes or even decreases slightly. This suggests that, after a certain threshold, increasing these parameters no longer significantly improves the performance of the algorithm. For population size, fitness continues to increase, but at a lower rate, which indicates a diminishing returns effect.

To quantify the strength of the relationships between the algorithm's adjustable parameters and fitness values, we also calculated Spearman correlation coefficients. The obtained values indicate a strong correlation: 0.918 for population size, 0.910 for the number of generations and 0.971 for the cooling factor.

In addition to the analysis performed by scatter plots and Spearman correlation coefficients, a regression model was applied to more precisely quantify the relationships between the adjustable parameters and the fitness value, as well as to evaluate the individual contribution of each parameter to the performance of the algorithm. For this purpose, a polynomial regression model of degree 2 was used, having as independent variables the population size, the number of generations and the cooling factor. Figure 9 shows the regression results.

```
                            OLS Regression Results
==============================================================================
Dep. Variable:                      y   R-squared:                       0.120
Model:                            OLS   Adj. R-squared:                  0.114
Method:                 Least Squares   F-statistic:                     19.23
Date:                Thu, 03 Apr 2025   Prob (F-statistic):           2.04e-30
Time:                        09:02:32   Log-Likelihood:                 5013.9
No. Observations:                1280   AIC:                         -1.001e+04
Df Residuals:                    1270   BIC:                            -9956.
Df Model:                           9
Covariance Type:            nonrobust
===============================================================================
                                 coef    std err          t      P>|t|      [0.025      0.975]
-------------------------------------------------------------------------------
Intercept                       0.8392      0.002    411.880      0.000       0.835       0.843
PopSize                         0.0001   2.04e-05      5.020      0.000    6.25e-05       0.000
GenerationNumber             6.425e-05   2.04e-05      3.142      0.002    2.41e-05       0.000
CoolingFactor                   0.0127      0.003      4.603      0.000       0.007       0.018
PopSize^2                    -1.889e-07   7.15e-08     -2.641      0.008   -3.29e-07   -4.86e-08
PopSize GenerationNumber     -1.615e-07   6.36e-08     -2.539      0.011   -2.86e-07   -3.67e-08
PopSize CoolingFactor        -1.312e-05   1.04e-05     -1.266      0.206   -3.34e-05    7.21e-06
GenerationNumber^2           -4.938e-08   7.15e-08     -0.691      0.490    -1.9e-07    9.09e-08
GenerationNumber CoolingFactor -2.594e-05  1.04e-05     -2.504      0.012   -4.63e-05   -5.61e-06
CoolingFactor^2                 -0.0076      0.002     -3.782      0.000      -0.012      -0.004
==============================================================================
Omnibus:                       15.703   Durbin-Watson:                   2.046
Prob(Omnibus):                  0.000   Jarque-Bera (JB):               25.044
Skew:                          -0.054   Prob(JB):                     3.65e-06
Kurtosis:                       3.677   Cond. No.                     8.12e+05
==============================================================================
```

Figure 7: Polynomial Regression Results

Based on obtained values we concluded that all three analysed parameters, population size (PopSize), number of generations (GenerationNumber) and cooling factor (CoolingFactor), have a significant influence on the fitness value. Among them, CoolingFactor has the largest positive impact. The quadratic terms associated with PopSize² and CoolingFactor² highlight the presence of diminishing returns effects, which means that, after a certain threshold, increasing these parameters no longer brings proportional improvements in fitness value. Also, the interactions between PopSize X GenerationNumber and GenerationNumber X CoolingFactor are associated with a negative effect on fitness, suggesting that a simultaneous increase in these pairs of variables may have an unfavourable impact. On the other hand, the terms PopSize X CoolingFactor and GenerationNumber² are not statistically significant, indicating that these factors do not significantly influence the fitness function. The regression model explains approximately 12% of the variability ($R^2$ = 0.12) in fitness values, indicating that a significant proportion of the variation in algorithm performance is determined by additional factors, which are not included in the current analysis. Regarding the interaction effects, the negative and significant coefficient for PopSize X GenerationNumber (coef. = -1.615e-07, p = 0.011) shows that a simultaneous increase in these parameters can have a negative impact on fitness. A similar effect is also observed for GenerationNumber X CoolingFactor (coef. = -2.594e-05, p = 0.012). In contrast, the PopSize X CoolingFactor interaction (coef. = -1.312e-05, p = 0.206) is not significant, suggesting that the combined influence of these two parameters does not have a noticeable effect on the performance of the algorithm.

To better evaluate the model performance and validate the regression hypotheses, we performed a detailed analysis of the residuals and checked the multicollinearity of the predictors. The distribution of residuals relative to the values predicted by the model plotted in Figure 8(a) shows that the residuals are generally randomly distributed around the zero axis. This means that the relationship between independent variables and fitness is approximately correctly modelled. However, a slight increase in the residuals dispersion is observed for lower values of predicted fitness, which could indicate a lack of homoscedasticity in this area. This suggests that the variance of errors is not constant over the entire range of predicted values and indicates the need for further adjustment of the model.

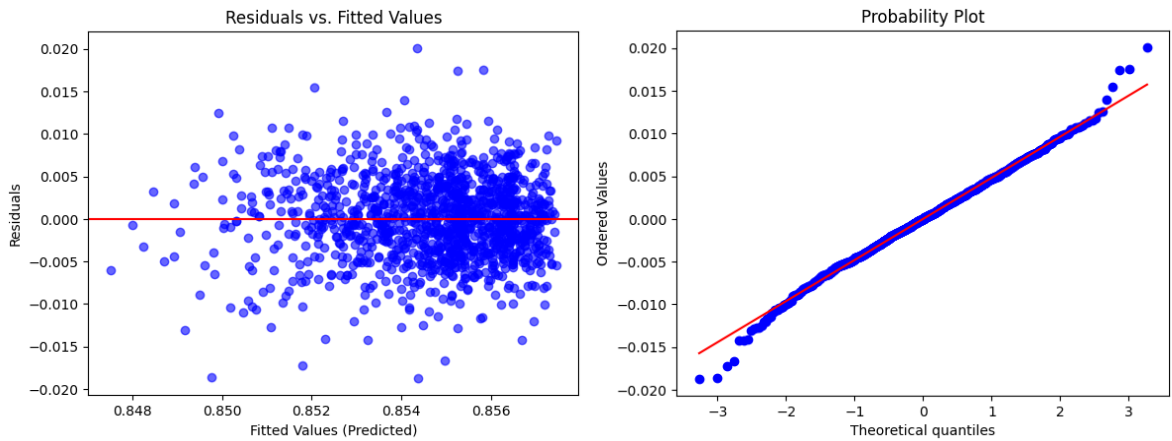

Figure 8: (a) Residuals vs. Predictor Plot scatter plot; (b) Q-Q Plot for residuals

The Q-Q (quantile-quantile) plot from Figure 8 (b) shows that that the distribution of the residuals is close to normal. Minor deviations from the diagonal line, visible at the lowest and highest quantiles, suggest the presence of outliers, but are not sufficiently pronounced to signal a significant violation of normality assumption. Therefore, the assumption of normality of the residuals is reasonably met, which supports the validity of the statistical significance tests applied within the regression model.

Figure 9 shows the variance inflation factor (VIF) values for the explanatory variables (i.e. number of generations, population size and cooling factor) included in the regression model. Values that slightly exceed the threshold of 5 for generation number and population size suggest a moderate collinearity, while the value below 5 for cooling factor suggest a low correlation with the other explanatory variables. All these values do not indicate major multicollinearity problems and can be considered acceptable in the context of our model.

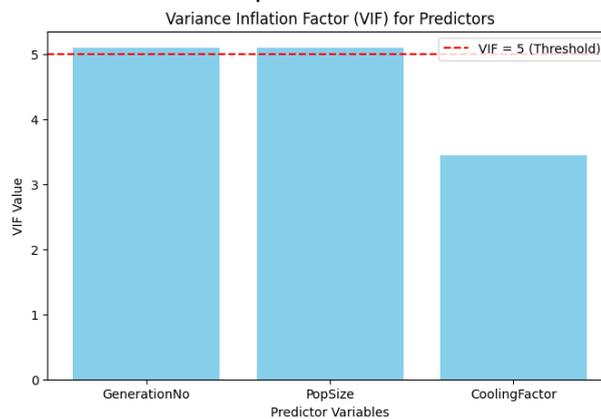

Figure 9: VIF for predictors

Finally, to further reduce the effects of multicollinearity and increase the stability of the coefficient estimates in the polynomial regression model, Ridge regularization was used, and the results obtained by applying it are shown in Figure 10. The value of the regularization hyperparameter α was set to 1.0 and was determined using Grid Search. The model performance was evaluated using the mean square error (MSE), and the small value obtained indicates a high accuracy of the predictions.

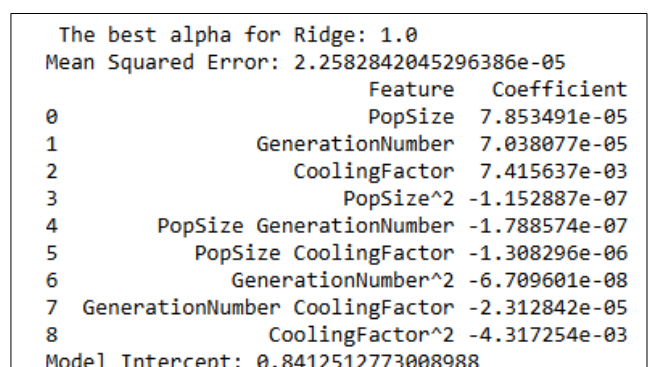

```
The best alpha for Ridge: 1.0
Mean Squared Error: 2.2582842045296386e-05
                       Feature     Coefficient
0                      PopSize    7.853491e-05
1               GenerationNumber  7.038077e-05
2                 CoolingFactor   7.415637e-03
3                      PopSize^2 -1.152887e-07
4       PopSize GenerationNumber -1.788574e-07
5          PopSize CoolingFactor -1.308296e-06
6              GenerationNumber^2 -6.709601e-08
7 GenerationNumber CoolingFactor -2.312842e-05
8                 CoolingFactor^2 -4.317254e-03
Model Intercept: 0.8412512773008988
```

Figure 10: Regression with Ridge regularization technique

The coefficients estimated by Ridge Regression shows that each parameter of the algorithm influences fitness differently. Of all, cooling factor has the largest positive effect, but this effect decreases for large values, according to the negative coefficient associated with the quadratic term, which indicates a saturation phenomenon. The population size and generation number also contribute to increasing fitness, but less, and in their case a decrease in impact is also observed at high values. The coefficients of the interaction suggest that the parameters do not act completely independently: the simultaneous increase of two parameters can produce an effect weaker than the sum of their individual effects. These interactions indicate a complex relationship between the parameters, which must be considered for efficient optimization of the algorithm's performance.

## 5.2. Algorithm Performance Evaluation for Optimal Configuration of adjustable parameters

Based on the detailed analysis of the relations between the adjustable parameters and the fitness value, we decided to explore various combinations of adjustable parameters to identify the one that maximizes the fitness value. The ranges for each adjustable parameter, was set as follows: [0.01, 1] for cooling factor, [50, 400] for population size, and [50,500] for generation number. Since the cooling factor has the greatest impact on the fitness value, we started with a small value and then increased it while the other two parameters were kept constant and analysed the impact it has on fitness. Similar experiments have performed for the other two parameters (i.e. population size and generation numbers). We also avoided combining large values for population size and number of generations or for number of generations and cooling factor, which would negatively affect the performance of the algorithm. Figure 11 shows the evolution of fitness over generations for the best six configurations of adjustable parameters.

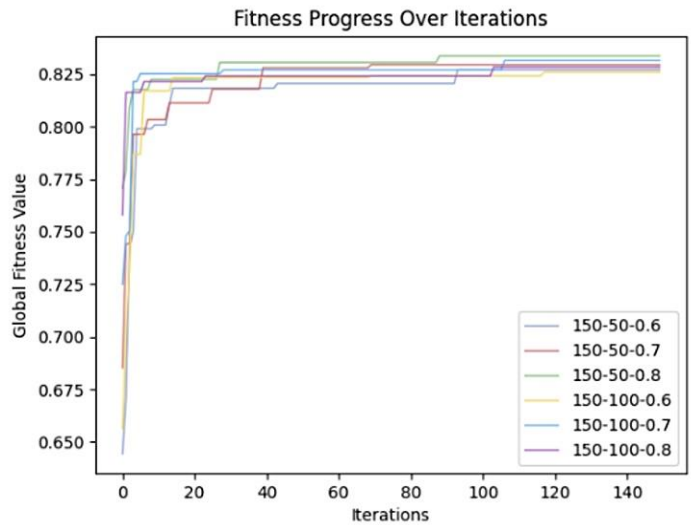

Figure 11: Fitness evolution over generations for different configurations of adjustable parameters

Analysing this graph, we can conclude that the best performing combination of adjustable parameters, according to the evolution of fitness, is 150-50-0.8, closely followed by the configurations 150-100-0.8 and 150-100-0.7. This indicates that a high number of generations and a cooling factor between 0.7 and 0.8 are essential for maximizing the performance of the algorithm in the tested configurations.

Figure 12 shows how population diversity varies across generations for the same configuration of adjustable parameters.

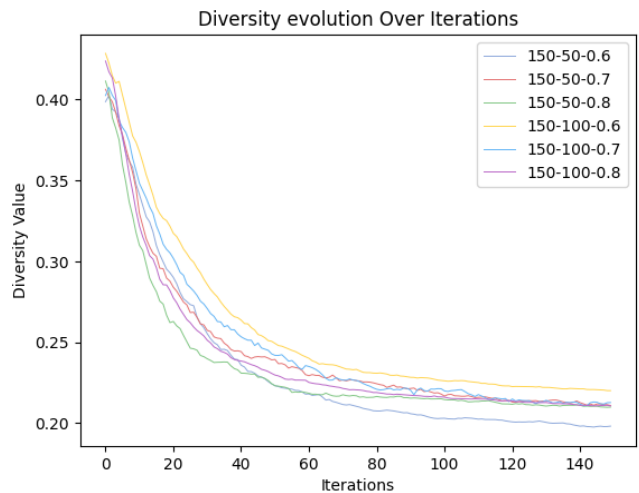

Figure 12: Diversity evolution over generations for different configurations of adjustable parameters

Analysis of the diversity evolution graph indicates that a larger population size, combined with a lower cooling factor, contributes to maintaining diversity throughout the iteration, providing a favourable framework for a more extensive exploration of the search space. In contrast, higher values of the cooling factor accelerate convergence, but may lead to a premature loss of diversity, increasing the risk of stagnation in a local optimum. Although the 150-50-0.8 configuration generated the best results in terms of fitness value, it is characterized by a rapid convergence and an early decrease in population diversity. Therefore, this configuration is recommended in situations where the main objective is to maximize performance in a short time. In contrast, the 150-100-0.7 configuration maintains a higher level of diversity over time and thus is more appropriate for problems with complex search spaces, in which maintaining the population diversity is essential to avoid premature stagnation.

Table 1: Execution time for different configurations of adjustable parameters

| Config | Execution Time(s) |
|---|---|
| generationNo =150, popSize = 50, coolingFactor = 0.6 | 3.9 |
| generationNo =150, popSize = 50, coolingFactor = 0.7 | 1.54 |
| generationNo =150, popSize = 50, coolingFactor = 0.8 | 1.87 |
| generationNo =150, popSize = 100, coolingFactor = 0.6 | 2.62 |
| generationNo =150, popSize = 100, coolingFactor = 0.7 | 4.1 |
| generationNo =150, popSize = 100, coolingFactor = 0.8 | 3.78 |

Based on this analysis we concluded that the configuration generationNo = 150, popSize = 50, coolingFactor = 0.8 represents the best choice for maximizing the performance of our algorithm. This configuration achieves the highest final fitness and fast convergence, with a short execution time (1.87 sec) (see Table 1), making it suitable for applications where computational efficiency and solution quality are priorities. Furthermore, although the level of diversity decreases relatively quickly in this configuration, this behaviour is acceptable in the context where the goal is to obtain a good result in a short time.

After identifying the optimal configuration of adjustable parameters, we performed a detailed analysis of the behaviour of the algorithm to better understand its performance over time. At this stage, we analysed several important aspects, including: the evolution of local and best fitness over generations, the stability of the algorithm between successive runs, the complexity of the algorithm, and the execution time. The stability of the algorithm was evaluated by analysing its performance under different random initialization conditions. Figure 13 shows the evolution of the fitness value over the iterations for five distinct runs of the algorithm, each starting from a different randomly generated initial population.

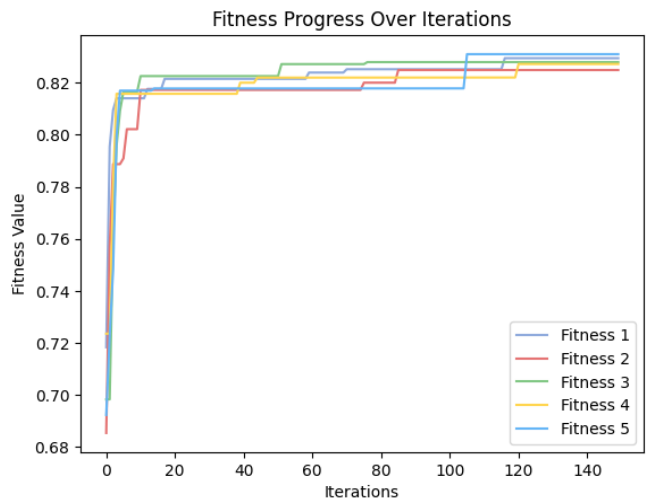

Figure 13: Fitness evolution from different randomly generated initial populations

The graph shows that all runs converge to similar fitness values, despite starting with randomly generated initial populations. This demonstrates the stability of the algorithm, namely that it produces similar results, even if the optimization process includes stochastic components.

# Conclusion

In this article, we proposed a method that combines a strategy based on game theory with an evolutive algorithm for identifying the optimal coalition of microgrids that trade directly with the energy market in the surplus or deficit energy periods. The goal is to maximize the profit from purchasing energy during periods of energy shortage in the markets and to minimize the costs associated with purchasing energy during periods of energy surplus in the market. In our approach, each microgrid is equipped with a storage unit, whose internal state is characterized by the available amount of energy, the number of remaining charge/discharge cycles, the maximum storage capacity, and the current operational state (i.e. charged, partially charged, or discharged). These attributes define the flexibility of the microgrid to participate in energy coalitions. The identification of the optimal coalition is achieved through an evolutionary algorithm and the distribution of benefit between the coalition members is achieved based on the Shapley value. The quality of the proposed approach was evaluated on the Zenodo dataset, and the results demonstrate that the algorithm is able to identify the optimal coalition capable of absorbing the market surplus. The microgrids selected to be part of the coalition are those that still have a large/medium available storage capacity and a large/medium number of remaining charging cycles.